\documentclass[twocolumn,showpacs,superscriptaddress]{revtex4}
\usepackage{amsmath}
\usepackage{amssymb}
\usepackage{bm}
\usepackage{epsfig}
\usepackage{graphicx}
\usepackage{color}
\usepackage[colorlinks=true]{hyperref}
\usepackage[english]{babel}

\begin{document}

\newcount\timehh  \newcount\timemm
\timehh=\time \divide\timehh by 60
\timemm=\time
\count255=\timehh\multiply\count255 by -60 \advance\timemm by \count255

\title{Observation of high angular momentum excitons in cuprous oxide}

\author{J. Thewes}
\author{J. Heck\"otter}
\author{T. Kazimierczuk}
\author{M. A\ss mann}
\author{D. Fr\"ohlich}
\affiliation{Experimentelle Physik 2, Technische Universit\"{a}t
Dortmund, D-44221 Dortmund, Germany}

\author{M. Bayer}
\affiliation{Experimentelle Physik 2, Technische Universit\"{a}t
Dortmund, D-44221 Dortmund, Germany} \affiliation{Ioffe Institute,
Russian Academy of Sciences, 194021, St.-Petersburg, Russia}

\author{M.A. Semina}
\author{M.M. Glazov}
\affiliation{Ioffe Institute, Russian Academy of Sciences, 194021,
St.-Petersburg, Russia}


\begin{abstract}
The recent observation of dipole-allowed $P$-excitons up to
principal quantum numbers of $n=25$ in cuprous oxide has given
insight into exciton states with unprecedented spectral resolution.
While so far the exciton description as a hydrogen-like complex has
been sufficient for cubic crystals, we demonstrate here distinct
deviations: The breaking of rotational symmetry leads to mixing of
high angular momentum $F$- and $H$-excitons with the $P$-excitons so
that they can be observed in absorption. The $F$-excitons show a
three-fold splitting that depends systematically on $n$, in
agreement with theoretical considerations. From detailed comparison
of experiment and theory we determine the cubic anisotropy parameter
of the Cu$_2$O valence band.
\end{abstract}

\maketitle

\emph{Introduction.} Excitonic effects are decisive for the optical
properties of semiconductors and insulators~\cite{exciton:gen}. Not
only leads the Coulomb interaction between an electron
and a hole to a series of bound states, the
\emph{excitons}, with energies below the band gap, but also above
the gap the Coulomb effects lead to a massive redistribution of
oscillator strength towards the low-energy states compared to a free
particle description. Due to this importance it has been a major
goal to develop a detailed understanding of excitons on a
quantitative level~\cite{exciton:gen}. The description of the bound
exciton states by the hydrogenic model has turned out to be
extremely successful in this respect, in particular, for bulk
semiconductors of cubic symmetry.

For excitons with wavefunction extensions much larger than the
crystal unit cell (the Mott-Wannier excitons) the hydrogen formula
for their binding energy, $\mathcal R/n^2$ with the Rydberg energy
$\mathcal R$ in a state of principal quantum number $n$, can be
simply adapted to the solid state case by (i) changing the reduced
mass of electron and proton $m$ to that of electron and hole $m^*$,
and (ii) screening the carrier interaction by the dielectric
constant $\varepsilon$: $\mathcal R^* = \mathcal R
m^*/(\varepsilon^2m)$. The influence of the many-body crystal
environment is thus comprised in material properties that for cubic
semiconductors are, as a rule, isotropic such as the scalar
dielectric constant $\varepsilon$, leading to a formula for
excitonic energies that is identical to the one in a system with
rotational symmetry. The material environment typically causes a
reduction of the atomic Rydberg energy by $2-3$ orders of magnitude
into the meV range.

For the hydrogen problem the spatial symmetry is determined by the
continuous rotation group SO(3), where the square of the orbital
momentum $L^2= l(l+1) \hbar^2$ and its $z$-component $L_z=m \hbar$,
the magnetic quantum number, are constants of motion, making the
problem integrable. Due to this symmetry the energy levels are
degenerate with respect to $m$. For the $1/r$-dependence of the
Coulomb potential, the Lenz-Runge vector is also conserved as
specific consequence of the underlying SO(4) symmetry, causing the
energy level degeneracy in $l$.

The latter degeneracy is lifted for hydrogen-like systems such as
Rydberg atoms~\cite{gallagher}, where one electron is excited into a
shell with $n \gg 1$. Here the screening of the nuclear Coulomb
potential by inner shell electrons causes a deviation from the $1/r$
behavior, breaking the SO(4) symmetry. Phenomenologically this can
be described by the quantum defect model, in which the binding
energy formula is modified to $\mathcal R/(n-\delta_l)^2$. The
quantum defect, $\delta_l$, depends on $l$, as the screening varies
with the angular momentum of the outmost electron state due to its
different penetration into the electron core. The $m$-degeneracy is
still maintained, though, for rotational symmetry.

In crystals this rotational symmetry is broken down to discrete
groups. For bulk cubic crystals these are the groups $O_h$ or $T_d$
(depending on presence or absence of inversion symmetry). Since both
groups represent still quite high symmetry, e.g., compared to
low-dimensional semiconductors, the deviations from the rotational
symmetry are usually captured solely by the lattice-periodic Bloch
functions of electron and hole. The envelope wave function of
Mott-Wannier excitons, however, are typically considered as
hydrogen-like involving the spherical harmonics as angular momentum
eigenfunctions, even though, strictly speaking, angular momentum is
no longer conserved, and $l,m$ are not good quantum numbers. So far,
no indications for a failure of this description such as observation
of a splitting of levels of particular $l$ and/or mixing of levels
with different $l$ have been reported.

Due to the small Rydberg energy in prototype semiconductors of
highest quality like GaAs ($\mathcal R^* = 4.2$ meV), consequences
of the reduced crystal symmetry are hard to resolve in optical
spectra of excitons. One might seek for deviations from the exciton
hydrogen model in materials with larger Rydberg energy such as
oxides or nitrides, for which possible exciton level splittings,
however, may be blurred by crystal inhomogeneities. A somewhat
unique position in terms of crystal quality is held by cuprous oxide
(Cu$_2$O) with a Rydberg energy of about 90 meV. The high quality of
Cu$_2$O natural crystals allowed the first experimental
demonstration of excitons~\cite{gross1,gross2}, and it is evidenced
also by remarkably narrow absorption lines as highlighted by the
observation of the paraexciton in magnetic field with a record
linewidth below 100 neV~\cite{brandt}.

Here we report a high resolution absorption study of the yellow
exciton series in cuprous oxide. Besides the dominant $P$-excitons
we resolve exciton triplets from $n=4$ to 10 with linewidths in the
$\mu$eV-range. These lines are ascribed to $F$-excitons (more
precisely, to mixed $F$- and $P$-excitons). Their emergence in the
spectra and their splitting which depends systematically on the
principle quantum number $n$ are consequences of the reduction from
full rotational to discrete $O_h$ symmetry. These findings are in
good agreement with calculations in which the band structure details
are taken into account.

\emph{Symmetry analysis.} The yellow exciton series is associated
with transitions between the highest valence and lowest conduction
bands in Cu$_2$O. In this material the topmost valence band
corresponds to the irreducible representation $\mathcal D_h =
\Gamma_7^+$ of the $O_h$ point group, and the bottom conduction band
corresponds to $\mathcal D_e = \Gamma_6^+$, see
Ref.~\cite{PhysRevB.23.2731} for details. The exciton wavefunction
transforms according to the product $\mathcal D_x=\mathcal D_e
\times \mathcal D_h \times \mathcal D_r$ where $\mathcal D_r$ is the
representation describing the wavefunction of the electron-hole
relative motion~\cite{ivchenko05a}.

To draw the analogy with hydrogen, it is convenient to follow the
conventional description and categorize the excitonic states by the
symmetry of the relative motion envelope $\mathcal D_r$. For
$S$-excitons and $P$-excitons $\mathcal D_r$ is irreducible,
$\mathcal D_S=\Gamma_1^+$ and $\mathcal D_P=\Gamma_4^-$,
respectively, while for the states with higher angular momentum,
e.g. the $F$ ($l=3$) and $H$ ($l=5$) excitons the representations
are reducible, $\mathcal D_F= \Gamma_2^- + \Gamma_4^- + \Gamma_5^-$
and $\mathcal D_H = \Gamma_3^-+2\Gamma_4^- + \Gamma_5^-$,
demonstrating that $l$ is no longer a good quantum number for cubic
symmetry. Still, we will maintain the exciton classification
according to orbital angular momentum for simplicity.

To be optically active, the exciton state representation $\mathcal
D_x$ has to contain the three-dimensional $\mathcal D_x =
\Gamma_4^-$ representation, corresponding to the components of the
electric dipole operator. The product of the electron and hole Bloch
functions can be decomposed as $\mathcal D_e \times \mathcal D_h =
\Gamma_2^+ + \Gamma_5^+$. Taking into account the envelope function,
the resulting $P$-excitons are optically active in one-photon
transitions because $\Gamma_4^-$ is contained in the product
representation: $\Gamma_4^- \in \mathcal D_P \times \Gamma_5^+$,
while this is not the case for the $S$- and $D$-excitons, so that
they are dark~\cite{Kazimierczuk:2014yq}. The origin of this
behavior is the even parity of conduction and valence bands in
Cu$_2$O so that dipole transitions have to involve odd envelopes.
However, accounting for the cubic crystal symmetry makes the $F$-
and $H$-excitons dipole allowed in addition to the $P$-excitons: One
can readily check that the product $\mathcal D_F \times \Gamma_5^+$
contains $3\Gamma_4^-$ and $\mathcal D_F \times \Gamma_2^+$ contains
one more $\Gamma_4^-$. Hence $F$-exciton states can give rise to
four spectral lines. However, one of those arising from the
$\Gamma_2^+$ product of the Bloch functions is weak ~\footnote{The
transitions to the $\Gamma_2^+$ state (paraexciton) are
spin-forbidden and can be activated by magnetic field~\cite{brandt,
gastev}. Owing to the cubic symmetry, the $\mathcal D_F\times
\Gamma_2^+$ state mixes with $\mathcal D_F \times \Gamma_5^+$ and
becomes allowed but remains weak, see Supplement for details.}.
Analogously, the product $\mathcal D_H \times
(\Gamma_2^++\Gamma_5^+)$ contains $5\Gamma_4^-$, hence $H$-exciton
can give rise to five lines.

\emph{Experimentals.} To test these predictions, high-resolution
absorption spectra were recorded by detecting the emission of a
frequency-stabilized laser with 1 neV linewidth (corresponding to
about 250 kHz band width) after transmission through a 30 $\mu$m
thick Cu$_2$O crystal slab. The wavelength of the laser emission was
scanned in the range of interest from 570 to 580 nm, for details see
Ref. \cite{Kazimierczuk:2014yq}. The sample was held in liquid
Helium at a temperature of 1.2~K and was strain-free mounted in a
holder that allowed also application of an electric field along the
optical axis.

\begin{figure}[t]
\includegraphics[width=\linewidth]{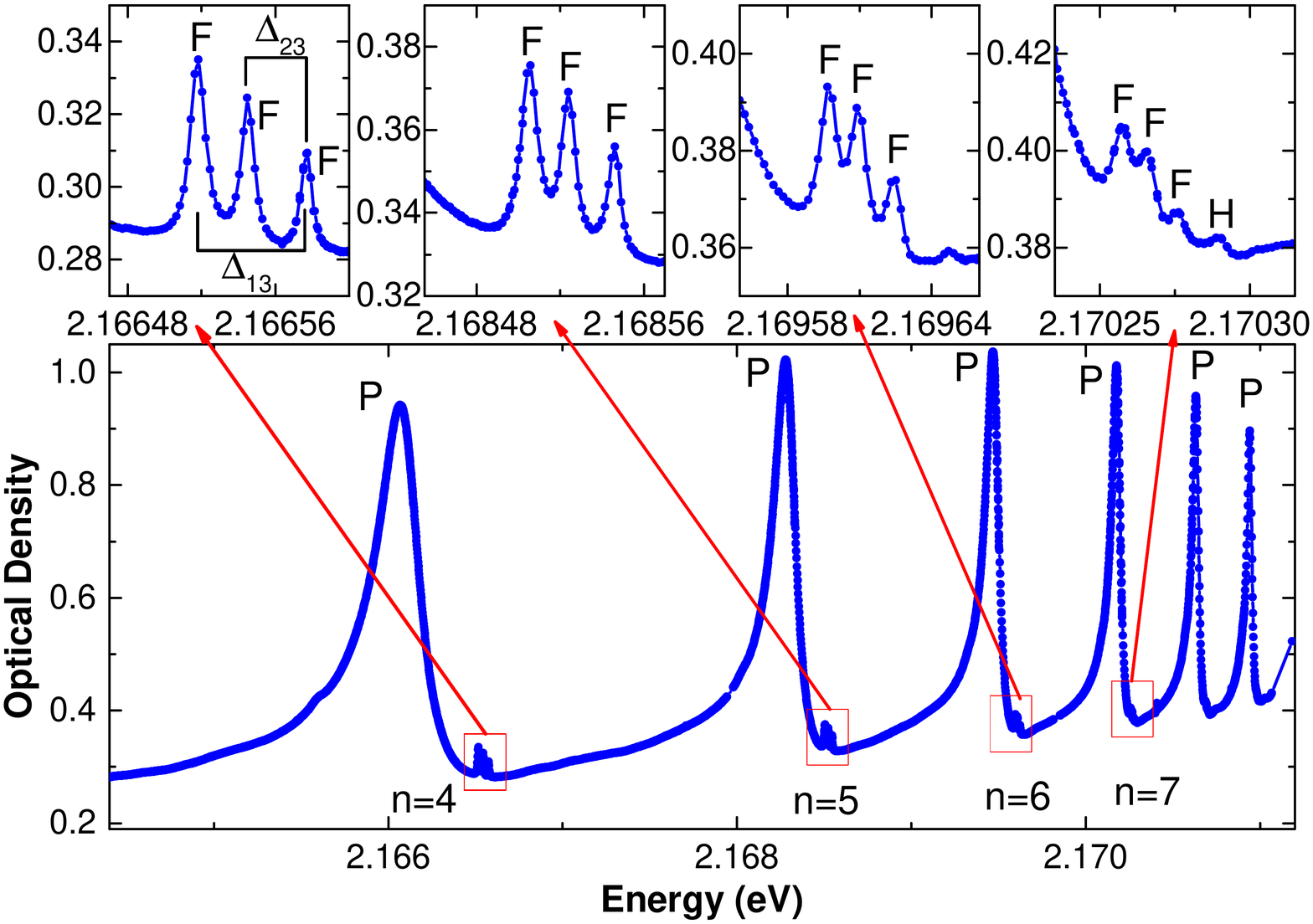}
\caption{Bottom: absorption spectrum of the Cu$_2$O yellow exciton
series in the energy range of states with principal quantum numbers
from $n=4$ to $n=9$. The top panels show close-ups of the high
energy flanks of the $P$-excitons with $n =4, 5, 6,$ and 7,
respectively. $T = 1.2$~K.} \label{fig:transmission}
\end{figure}

The bottom panel of Fig.~\ref{fig:transmission} shows an absorption
spectrum of the yellow exciton series in the energy range
corresponding to principal quantum numbers from $n=4$ to $n=9$. The
spectrum is dominated by strong absorption features of the
$P$-excitons, discussed in detail in
Ref.~\cite{Kazimierczuk:2014yq}. However, on the high energy flank
of these lines weak additional features appear. We highlight that
these features can be observed starting from $n=4$ only. The top
panels show close-ups of the high energy flanks of the $n =4, 5, 6,$
and 7 $P$-excitons. Each group of features consists of a triplet of
lines with the splitting between them decreasing systematically with
increasing $n$. The width of each line is in the $\mu$eV-range, also
decreasing with $n$ which corresponds to lifetimes in the nanosecond
range. Starting from $n=6$ another feature appears on the high
energy side of the triplet, approaching the triplet with increasing
principal quantum number.

\begin{figure}[t]
\includegraphics[width=\linewidth]{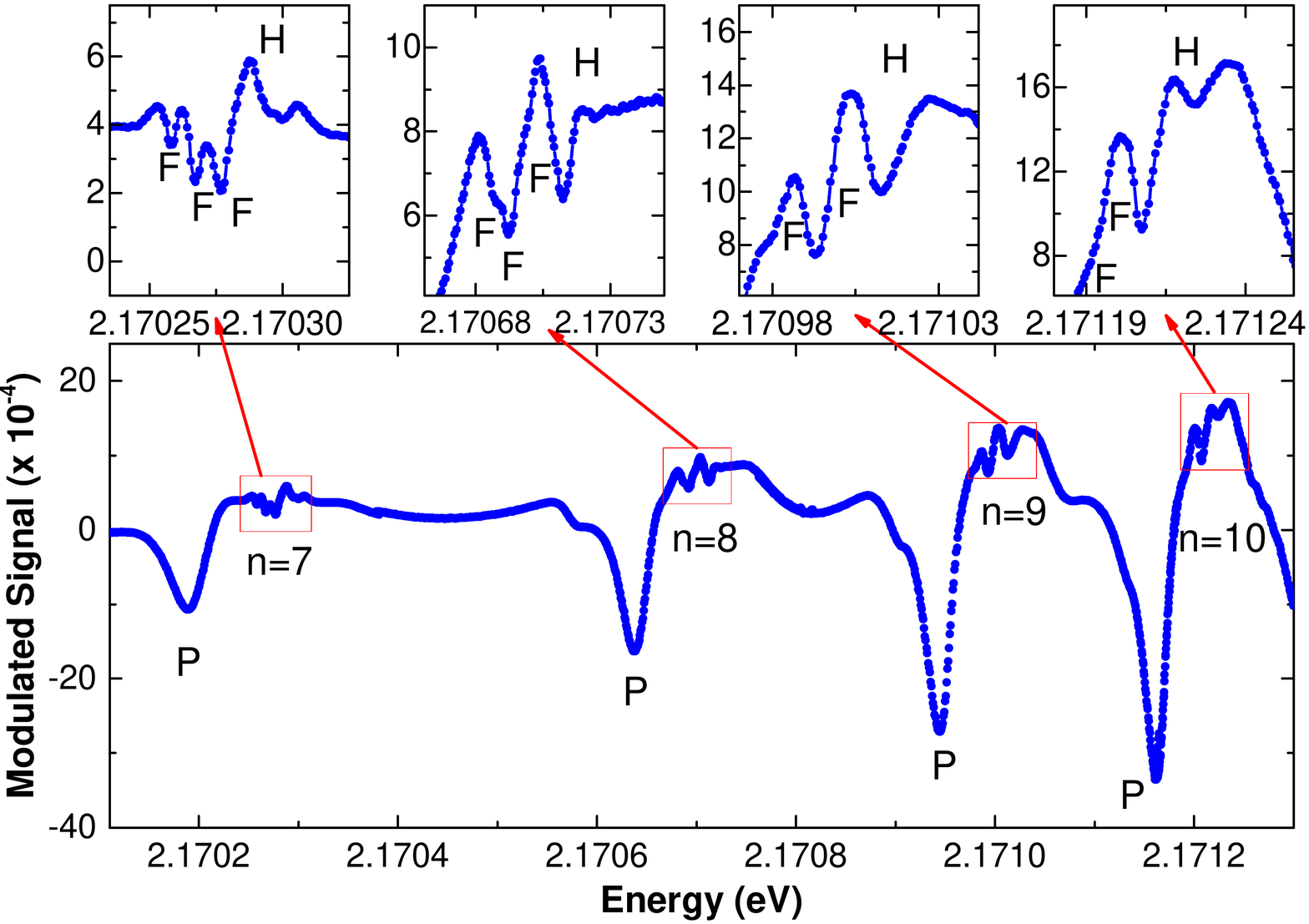}
\caption{Modulation spectra recorded as difference of transmission
spectra with and without a small electric field applied. The
electric field strength is 15 V/cm. The bottom panel shows the
spectral range of excitons from $n=7$ up to $10$; the top panels
give close-ups of the corresponding $F$-exciton features.}
\label{fig:transmission:mod}
\end{figure}

For $n>7$ the triplet is too close to the $P$-excitons to be
resolved in absolute transmission. Therefore we have applied
modulation spectroscopy by recording the differential absorption
with and without an electric field applied. Only a small field of
15V/cm is applied to avoid notable modifications both of absolute
exciton energies as well as splittings between them. The bottom
panel in Fig.~\ref{fig:transmission:mod} shows such a modulation
spectrum, in which triplets can be seen up to higher principal
quantum numbers. For comparison with Fig.~\ref{fig:transmission},
the energy range around the $n=7$ exciton is shown, where one
recognizes the triplet at the same energies as before. For $n=8$,
also the triplet can be seen, where however the two low energy lines
have almost merged, while the high energy line is still well
separated from them. For higher $n$ only a doublet of lines can be
seen, as the splitting between the low energy lines is comparable or
smaller than their line widths.

The angular momentum of exciton states belonging to a particular $n$
is limited by $n-1$, hence, in the hydrogen model the $F$-states
appear only for $n\geqslant 4$. Since the observed triplet indeed
emerges in ~Fig.~\ref{fig:transmission} starting from $n=4$ only, we
therefore assign it to $F$-excitons. We note that based on the
symmetry considerations presented above, $F$-excitons belonging to a
particular $n$ are expected to appear in optical spectra as a
quadruplet (with one line being weak). The numerical calculations
presented below confirm this and demonstrate that one of the lines
is indeed substantially weaker as compared with three others.
Moreover, its energy is very close to that of one strong line. For
additional confirmation, experiments in moderate magnetic fields
(not shown here) were carried out that demonstrate a splitting into
$6$ lines, in agreement with our symmetry analysis which predicts a
two-fold splitting of each line in the triplet. Furthermore, the
additional feature emerging from $n=6$ onwards can be attributed to
$H$-excitons. Their expected splitting is, however, too small to be
resolved.

To the best of our knowledge, such high angular momentum exciton
states including their fine structure splitting have not been
resolved so far. We emphasize here that the description as a
hydrogen-like complex can explain neither the optical activity of
$F$- and $H$-excitons, nor their splitting. It is therefore a unique
signature of the breaking of the rotational symmetry in the cubic
crystal and a consequence of its discrete symmetry. For a detailed
understanding we have developed a microscopic description of the
$F$-exciton fine structure.

\emph{Microscopic theory.} In Cu$_2$O the excitonic Rydberg
$\mathcal R^* \approx 90$~meV and the splitting between the
$\Gamma_7^+$ and $\Gamma_8^+$ valence bands $\Delta\approx 130$~meV
have the same order of magnitude. Therefore, a consistent theory of
excitonic states has to be based on treating the complex valence
band structure and the Coulomb interaction on the same level.
Correspondingly, we follow the approach of
Ref.~\cite{PhysRevB.23.2731} and present the exciton Hamiltonian
(for zero center-of-mass wavevector) in the form
\begin{equation}
\label{H} \mathcal H = \frac{p^2}{\hbar^2} - \frac{2}{r} -
\frac{\mu}{3\hbar^2} \left( P^{(2)} \cdot I^{(2)} \right) +
\frac{2}{3} \bar \Delta (1+ \bm I \cdot \bm s_h) + \mathcal H_c.
\end{equation}
Here $\bm p$ is the momentum of the relative electron-hole motion,
$\bm I$ is the angular momentum one operator acting in the basis of
the orbital hole Bloch functions $\Gamma_5^+$, and $\bm s_h$ is the
hole spin operator ($s_h=1/2$). The energies are measured in units
of the ``bulk'' excitonic Rydberg $\mathcal R^* = e^4
m_0/(2\hbar^2\varepsilon^2\gamma_1')$, the distances are measured in
units of the corresponding Bohr radius, $a^* = \hbar^2 \varepsilon
\gamma_1'/(e^2m_0)$, $\varepsilon$ is the static dielectric
constant, $\bar \Delta=\Delta/\mathcal R^*$ is the dimensionless
splitting between the $\Gamma_7^+$ and $\Gamma_8^+$ bands,
$\gamma_1' = \gamma_1 + m_0/m_e$, $\mu = {(}6\gamma_3 +
4\gamma_2{)}/({5\gamma_1'})$, $m_e$ is the conduction electron mass,
and the $\gamma_i$ ($i=1,2,3$) are the Luttinger parameters. In
contrast to Ref.~\cite{PhysRevB.23.2731} we include into the
Hamiltonian~\eqref{H} the contribution of the cubic symmetry
responsible for the valence band warping~\cite{PhysRevB.9.1525}
\begin{equation}
\label{H:cubic}
\mathcal H_c = \frac{\delta}{3\hbar^2} \left(\sum_{k=\pm 4}[P^{(2)} \times I^{(2)}]_k^{(4)}
+ \frac{\sqrt{70}}{5} [P^{(2)} \times I^{(2)}]_0^{(4)}\right),
\end{equation}
with $\delta = (\gamma_3 -\gamma_2)/\gamma_1'$, This extension is
crucial, as our calculations show, to describe the fine structure of
$F$-exciton states absent otherwise. The central-cell correction to
the Coulomb potential as well as the short-range electron-hole
exchange interaction are disregarded since for $P$- and $F$-excitons
the wave function of the relative motion vanishes for coinciding
electron and hole coordinates. In Eqs.~\eqref{H} and \eqref{H:cubic}
we use $P^{(2)}$ and $I^{(2)}$ for the second-rank irreducible
components of the tensors $p_ip_j$ and $I_i I_j$, where $i,j=x,y,z$
and $p_i$, $I_i$ are the Cartesian components of $\bm p$ and $\bm
I$, respectively. We note that the quartic terms in the dispersion,
like $p_x^4+p_y^4+p_y^4$, which are allowed in $O_h$, result in a
$F$-$P$-mixing and make the $F$-states active but do not cause their
splitting.

\begin{figure}[t]
\includegraphics[width=\linewidth]{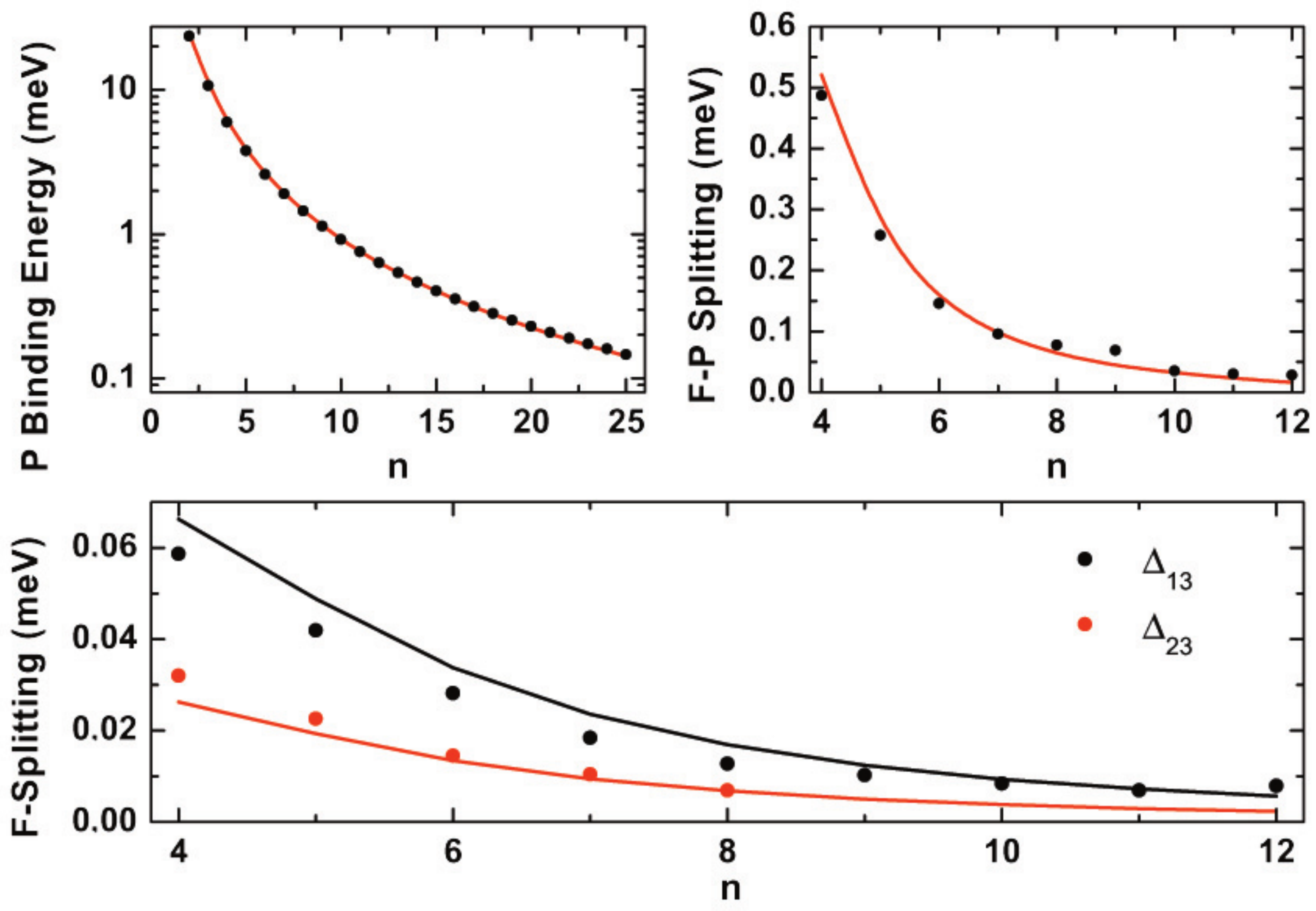}
\caption{Comparison of theoretical calculations and experimental
data for $P$-exciton binding energy (a), splitting between $F$- and
$P$-excitons (b), and splittings of $F$-triplet as defined in Fig.~1
(c) {\sl vs} principal quantum number. Dots give experimental data.
Theory is shown by lines connecting calculated data for discrete
values of $n$. The parameters of the calculation are: $\mathcal
R^*=87$~meV, $\mu=0.47$, $\Delta=134$~meV~\cite{PhysRevB.23.2731},
and $\delta=-0.1$.} \label{fig:compar}
\end{figure}

The Hamiltonian~\eqref{H} without cubic contribution $\mathcal H_c$
($\delta=0$) has full rotational symmetry and already provides a
quite accurate description of the $P$-exciton state
energies~\cite{PhysRevB.23.2731}. Therefore, it is instructive to
disregard $\mathcal H_c$ and determine the spectrum and
wavefunctions of $P$- and $F$-excitons. Moreover, the suppressed
short-range electron-hole exchange interaction makes it possible to
disregard the electron spin and characterize the excitons by the
hole total momentum $F$ and its $z$-component $F_z$, where $\bm F =
\bm J + \bm L$ is the sum of the momentum of the hole Bloch
functions $\bm J = \bm s_h + \bm I$ and the envelope function
orbital momentum $\bm L$.

The $P$-envelopes of excitons correspond to $F=1/2$ and $F=3/2$
states~\cite{PhysRevB.23.2731}, while $F$-excitons correspond in
this approximation to $F=5/2$ and $7/2$, respectively. Making use of
its symmetry one can express the wavefunction of a state with given
$F$ via basic functions with total momentum $F$, $|LJF\rangle$, and
radial functions $g_{LJ}(r)$, $\Psi(\bm r) = \sum_{LJ}
g_{LJ}(r)|LJF\rangle$ (see
Refs.~\cite{baldereschi73,PhysRevB.23.2731} for details), where the
functions $g_{LJ}(r)$ are determined numerically following the
method described in Refs.~\cite{baldereschi73,Semina:2011aa}, see
Supplement~\cite{Suppl} for details.

With inclusion of the cubic symmetry, $F$ and $F_z$ are no longer
good quantum numbers. In particular, the $F=5/2$ state gives rise to
$\Gamma_{6}^-$ and $\Gamma_8^-$ states while $F=7/2$ gives rise to
$\Gamma_6^-$, $\Gamma_7^-$ and $\Gamma_8^-$ states. Note, that
$\Gamma_6^-$ and $\Gamma_8^-$ are optically active ($\Gamma_4^-\in
\Gamma_6^-\times \mathcal D_e$, $\Gamma_8^-\times \mathcal D_e$),
while the $\Gamma_7^-$ states are dark. States of the same symmetry
have to be treated as close-to-degenerate, see
Refs.~\cite{PhysRevB.9.1525} and \cite{Suppl} for details. We have
calculated the energy spectrum of the $P$- and $F$-excitons for
different values of the cubic anisotropy parameter $\delta$ keeping
all other parameters known from literature fixed. Our calculations
show that the reasonably small value of $\delta = -0.1$ gives good
accord with experiment\cite{Suppl}. In this case two optically
active states out of four have very close energies. Moreover, one of
those states, namely, $\Gamma_6^-$ originating from $F=7/2$ has
small oscillator strength\cite{Suppl}. Smaller/larger values of
$\delta$ result in too small/large predictions for the $F$-shell
splitting, so that the estimate of $\delta$ from the experimental
data is quite accurate.

\emph{Discussion.} Figures~\ref{fig:compar}(a) and (b) compare the
results of calculations of the $P$-exciton binding energy and the
splitting between $F$- and $P$- excitons as function of principal
quantum number $n$ (the solid lines connect calculated values for
discrete $n$) with the measured data (dots), and serve as reference
for the accuracy of the description of the exciton states by our
model. For the $F$-excitons we have calculated the center-of-gravity
of the $F$-exciton lines. The comparison shows excellent agreement
between theory and experiment, providing confidence that also the
$\mu$eV splittings between the $F$-excitons can be assessed by our
model.

These splittings are shown in Fig.~\ref{fig:compar}(c), where
$\Delta_{13}$ is the energy separation between the outermost levels
and $\Delta_{23}$ is the energy separation between the middle and
higher energy level, see Fig.~1. Note, that the calculated energy of
the fourth (weakly) active state coincides within the accuracy of
several percent with the energy of middle state in the triplet. The
points with $4\leqslant n \leqslant 7$ were measured in absolute
transmission (see Fig.~\ref{fig:transmission}), while points with $n
\geq 8$ were taken from modulation spectroscopy. Both magnitudes of
the splittings and their dependence on $n$ are in good agreement
with our calculations. The splitting between the states decreases
strongly with $n$, as higher-$n$ states are more extended in space
so that their wave function averages over more crystal unit cells
and become less sensitive to crystal symmetry deviations from the
full rotation group. More quantitatively, the decrease is due to the
fact that the states are intermixed by quadratic combinations of the
momentum operator $p_i p_j$. The corresponding interaction matrix
elements scale as the inverse square of the state radii. In the
hydrogen model $\langle r^{-2}\rangle \propto n^{-3}(l+1)^{-1}$,
indicating that the splittings for higher angular momentum states
decrease with $n$ and $l$. This explains the missing splitting of
the $H$-excitons in absorption.

The agreement of the developed model with the experimental data
allows us to accurately evaluate the valence band parameters in
Cu$_2$O. Taking the static dielectric constant
$\varepsilon=7.5$~\cite{Land}, the electron effective mass
$m_e=0.99m_0$~\cite{mass}, and using our values for $\mathcal R^*$,
$\mu$ and $\delta$ (see text above and caption to
Fig.~\ref{fig:compar}), we obtain the following values of the
Luttinger parameters: $\gamma_1 =1.79$, $\gamma_2=0.82$, and
$\gamma_3=0.54$. Interestingly, in this material
$\gamma_3<\gamma_2$, which is also indicated by microscopic
calculations~\cite{models}, see Supplement. We emphasize that the
obtained Luttinger parameters may serve as benchmark for theoretical
models of the Cu$_2$O band structure.

\emph{Conclusion.} In summary, we have discovered high angular
momentum exciton states in one-photon absorption spectra in
high-quality Cu$_2$O bulk crystals. Observation of these states
becomes possible through the rotational symmetry breakdown by the
cubic crystal environment. Even though this symmetry breaking is
weak due to the high $O_h$ symmetry, the resulting level splittings
could be resolved because of the for crystals record-low linewidths
of the absorption features.

\emph{Acknowledgements.} We acknowledge the support by the Deutsche
Forschungsgemeinschaft and the Russian Foundation for Basic Research
in the frame of ICRC TRR 160, the RF President grants
MD-5726.2015.2, NSh-5062.2014.2 and NSh-1085.2014.2, Dynasty
Foundation, and programs of RAS.


\newpage

\section*{SUPPLEMENTARY INFORMATION}

\section{Symmetry analysis}

In the $O_h$ point group the $P$-shell excitonic states (quantum
number $l=1$) transform according to the irreducible representation
$\Gamma_4^-$ (hereinafter we use the notations of
Ref.~\cite{koster63}) whose basic functions can be chosen as $x$,
$y$ and $z$. The representation $\mathcal D_3^-$ ($F$-shell states,
$l=3$) of the full spherical symmetry group, $\mathcal K_h$, is
reducible in $O_h$ and can be decomposed into the following three
irreducible representations:
\begin{equation}
\label{D3m:decomp}
\mathcal D_{3}^{-} = \Gamma_2^- + \Gamma_4^- + \Gamma_5^-.
\end{equation}
Note that in Ref.~\cite{koster63} the compatibility table of
$\mathcal K_h$ is given for the $T_d$ group, where $\mathcal
D_3^-=\Gamma_1+\Gamma_4+\Gamma_5$, so the following replacements are
needed $\Gamma_1~(T_d)\to \Gamma_2^-~(O_h)$, $\Gamma_4~(T_d) \to
\Gamma_5^-~(O_h)$ and $\Gamma_5~(T_d) \to \Gamma_4^-~(O_h)$
according to the compatibility of $T_d$ and $O_h$. As follows from
Eq.~\eqref{D3m:decomp} the $7$-fold degenerate $F$-shell excitonic
state splits into the singlet $\Gamma_2^-$ [basic function $\propto
xyz\propto Y_{3,2}(\vartheta,\varphi) - Y_{3,-2}(\vartheta,\varphi)$
(where $\vartheta$ and $\varphi$ are the angles of $\bm r$ in
spherical coordinates), and two triplets $\Gamma_4^-$ [basic
functions transform like $x^3$, $y^3$, $z^3$, particularly, $x^3+
\mathrm i y^3 \propto \sqrt{5} Y_{3,-3}(\vartheta,\varphi) +
\sqrt{3} Y_{3,1}(\vartheta,\varphi)$, $x^3- \mathrm i y^3 \propto
-\sqrt{5} Y_{3,3}(\vartheta,\varphi) - \sqrt{3}
Y_{3,-1}(\vartheta,\varphi)$, $z^3 \propto
Y_{30}(\vartheta,\varphi)$] and $\Gamma_5^-$ [basic functions like
$x(y^2-z^2)$, $y(z^2-x^2)$, $z(x^2-y^2)$, where, correspondingly,
$x(y^2-z^2) + \mathrm i y(z^2 -x^2) \propto \sqrt{3}
Y_{3,3}(\vartheta,\varphi) - \sqrt{5} Y_{3,-1}(\vartheta,\varphi)$,
$x(y^2-z^2) - \mathrm i y(z^2 -x^2) \propto -\sqrt{3}
Y_{3,-3}(\vartheta,\varphi) + \sqrt{5} Y_{3,1}(\vartheta,\varphi)$,
$z(x^2-y^2) \propto Y_{3,2}(\vartheta,\varphi)+
Y_{3,-2}(\vartheta,\varphi)$].

With allowance for the spin and spin-orbit coupling the topmost
valence band transforms according to the $\Gamma_7^+$ representation
and the lowest conduction band transforms according to the
$\Gamma_6^+$ representation of the $O_h$ point symmetry group, see
Ref.~\cite{PhysRevB.23.2731} for details. For the $F$-shell envelope
function one has
\begin{multline}
\label{f:shell:full} \mathcal D_{3}^{-} \times \Gamma_7^+ \times
\Gamma_6^+ = (\Gamma_2^- + \Gamma_4^- + \Gamma_5^-)\times
(\Gamma_2^+ + \Gamma_5^+) \\= 2\Gamma_1^- + \Gamma_2^- + 2\Gamma_3^-
+ 4 \Gamma_4^- +3 \Gamma_5^-.
\end{multline}
Note, that $\Gamma_4^-$ appears 4-times, one of these
representations arises from the $\Gamma_2^+$ combination of electron
and hole Bloch functions, namely,
\[
|\Gamma_2^+\rangle = -\frac{1}{\sqrt{2}} |\Gamma_6^+,-1/2\rangle |\Gamma_7^+,1/2\rangle
\]
\[
+ \frac{1}{\sqrt{2}} |\Gamma_6^+,1/2\rangle |\Gamma_7^+,-1/2\rangle,
\]
and $\Gamma_4^- = \Gamma_5^-\times \Gamma_2^+$ with basic
combinations $x(y^2-z^2)|\Gamma_2^+\rangle$,
$y(z^2-x^2)|\Gamma_2^+\rangle$ and $z(x^2-y^2)|\Gamma_2^+\rangle$.
The $\Gamma_2^+$ representation corresponds to the paraexciton Bloch
function, the transition to these $\Gamma_4^-$ states is
spin-forbidden, unless those states are mixed with other states of
$\Gamma_4^-$ symmetry.

In numerical calculations (see main text for details) it is
convenient to consider the representation of the product of the hole
Bloch function and the excitonic envelope, in which case we have
\begin{equation}
\label{hole:sum} \mathcal D_{3}^{-} \times \Gamma_7^+  = (\Gamma_2^-
+ \Gamma_4^- + \Gamma_5^-)\times \Gamma_7^+ = 2\Gamma_6^- +
\Gamma_7^- + 2\Gamma_8^-,
\end{equation}
particularly,
\[
\Gamma_2^-\times \Gamma_7^+ = \Gamma_6^-, \quad \Gamma_4^-\times \Gamma_7^+ = \Gamma_7^- + \Gamma_8^- , \quad \Gamma_5^- \times \Gamma_7^+ = \Gamma_6^- + \Gamma_8^-.
\]
It is instructive to introduce the total momentum $\bm F$ of the
hole state, where neglecting cubic symmetry effects, states with
$F=5/2$ and $F=7/2$ are possible for the $F$-shell excitons. The
total angular momentum is not a good quantum number in the $O_h$
point group, and the corresponding representations are
reducible~\footnote{In derivation of Eqs.~\eqref{52sym} and
\eqref{72sym} one has to take into account that the hole spin
representation is $\Gamma_7^+$ (while a real spin $1/2$ transforms
according to $\Gamma_6^+$.}:
\begin{subequations}
\begin{equation}
\label{52sym}
\mathcal{ \tilde D}^-_{5/2} = \Gamma_6^- + \Gamma_8^-,
\end{equation}
whose basic functions are (defined as in Ref.~\cite{abragam1970electron}):
\begin{equation}
\label{basic52:7}
\Gamma_6^-: \quad \frac{1}{\sqrt{6}}|5/2\rangle - \sqrt{\frac{5}{6}}|-3/2\rangle,   |\frac{1}{\sqrt{6}}|-5/2\rangle - \sqrt{\frac{5}{6}}|3/2\rangle,
\end{equation}
\begin{multline}
\label{basic52:8}
\Gamma_8^-: \quad -\frac{1}{\sqrt{6}}|3/2\rangle - \sqrt{\frac{5}{6}}|-5/2\rangle, \\ |1/2\rangle, \quad -|-1/2\rangle, \\ \frac{1}{\sqrt{6}}|-3/2\rangle + \sqrt{\frac{5}{6}}|5/2\rangle,
\end{multline}
\end{subequations}
and
\begin{subequations}
\label{72sym}
\begin{equation}
\mathcal{\tilde D}^-_{7/2} = \Gamma_6^-+ \Gamma_7^- + \Gamma_8^-,
\end{equation}
with basic functions:
\begin{equation}
\label{basic:ABC}
\Gamma_6^-: \quad \frac{\sqrt{3}}{2} |5/2\rangle - \frac{1}{2} |-3/2\rangle, \quad \frac{1}{2} |3/2\rangle -\frac{\sqrt{3}}{2}|-5/2\rangle,
\end{equation}
\begin{multline}
\label{G7-}
\Gamma_7^-:  \quad \sqrt{\frac{5}{12}}|-7/2\rangle + \sqrt{\frac{7}{12}}|1/2\rangle, \\ -\sqrt{\frac{5}{12}}|7/2\rangle - \sqrt{\frac{7}{12}}|-1/2\rangle,
\end{multline}
\begin{equation}
\Gamma_8^-: \quad \frac{\sqrt{3}}{2} |3/2\rangle +\frac{1}{2}|-5/2\rangle,
\quad  \sqrt{\frac{7}{12}}|-7/2\rangle - \sqrt{\frac{5}{12}}|1/2\rangle,
\end{equation}
\[
\sqrt{\frac{7}{12}}|7/2\rangle - \sqrt{\frac{5}{12}}|-1/2\rangle,\quad \frac{1}{2} |5/2\rangle + \frac{\sqrt{3}}{2} |-3/2\rangle.
\]
\end{subequations}
The numbers in the ``kets'' denote the projections of $F_z$. Making
use of Eq.~\eqref{hole:sum} one obtains the optically active states
$\Gamma_4^- \in \Gamma_6^-\times \Gamma_6^+$ and $\Gamma_4^- \in
\Gamma_8^-\times \Gamma_6^+$.

The selection rules for optical transitions from the valence band
states Eqs.~\eqref{52sym}, \eqref{72sym} are similar to the
transitions in GaAs-like semiconductors ($T_d$ point symmetry) with
the replacement $\Gamma_7 (T_d) \leftrightarrow\Gamma_6^- (O_h)$.

\section{Computational approach}

To obtain the fine structure of the $F$-shell exciton states we
follow the approach of Refs.~\cite{PhysRevB.23.2731,baldereschi73}.
We start in the spherical approximation, which takes into account
explicitly the mixing of the $\Gamma_7^+$ and $\Gamma_8^+$ valence
subbands and afterwards we allow for cubic contributions.

\subsection{Spherical approximation}

Furthermore, we neglect the central-cell corrections to the
electron-hole potential and the electron-hole exchange interaction.
The Hamiltonian for the exciton with center of mass momentum $\bm
P=0$ has the form [cf. Eq. (1) of the main text]:
\begin{equation}
\label{H}
\mathcal H = \frac{p^2}{\hbar^2} - \frac{2}{r} - \frac{\mu}{3\hbar^2} \left( P^{(2)} \cdot I^{(2)} \right) + \frac{2}{3} \bar \Delta (1+ \bm I \cdot \bm s_h),
\end{equation}
where $p$ is the momentum of relative electron-hole motion, $\bm I$
is the angular momentum one operator acting in the basis of orbital
hole Bloch functions $\Gamma_5^+$, $\bm s_h$ is the hole spin
operator ($s_h=1/2$), the energies are measured in units of the
``bulk'' excitonic Rydberg $\mathcal R^* = e^4
m_0/(2\hbar^2\varepsilon^2\gamma_1')$, the distances are measured in
units of the corresponding Bohr radius, $a_0 = \hbar^2 \varepsilon
\gamma_1'/(e^2m_0)$, $\varepsilon$ is the static dielectric
constant, $\bar \Delta=\Delta/R_0$ is the dimensionless splitting
between the $\Gamma_7^+$ and $\Gamma_8^+$ bands, $\gamma_1' =
\gamma_1 + m_0/m_e$,
\[
\mu = \frac{6\gamma_3 + 4\gamma_2}{5\gamma_1'},
\]
and $\gamma_i$ ($i=1,2,3$) are the Luttinger parameters.

In the spherical approximation the good quantum numbers are $F$ and
$F_z$, where $\bm F = \bm L + \bm J$ is the total angular momentum
of the envelope, $\bm L$, and the hole, $\bm J=\bm I + \bm s_h$,
states. The wavefunction is written as
\begin{multline}
\label{Psi:gen}
\Psi(\bm \rho) = \sum_{LJ} g_{LJ}(\rho) |LJF\rangle\\ = \sum_{LL_z,JJ_z} g_{LJ}(\rho) Y_{LL_z}(\theta,\varphi) C^{FF_z}_{LL_z;JJ_z} |J,J_z\rangle,
\end{multline}
where $ g_{LJ}(\rho)$ are the radial functions, $Y_{LL_z}$ are the
spherical harmonics, $C^{FF_z}_{LL_z;JJ_z}$ are the Clebsch-Gordan
coefficients, and $|J,J_z\rangle$ are the basic functions of the
valence band top. The effective Hamiltonian acting on the radial
functions can be constructed making use of the Appendices of
Refs.~\cite{PhysRevB.23.2731,baldereschi73}. Its matrix elements
between the $|LJF\rangle$ and $|L'J'F'\rangle$ states read
\begin{widetext}
\begin{subequations}
\label{Hamiltonian}
\begin{equation}
\langle LJF| \left(\frac{p^2}{\hbar^2} - \frac{2}{r} \right)|L'J'F'\rangle = \delta_{LL'}\delta_{JJ'}\delta_{FF'} \left(-\frac{d^2}{dr^2} - \frac{2}{r} \frac{d}{dr} + \frac{L(L+1)}{r^2} - \frac{2}{r} \right),
\end{equation}
\begin{equation}
\label{Hmu}
\langle LJF| \left(P^{(2)} \cdot I^{(2)} \right )|L'J'F'\rangle = 3\sqrt{5}\delta_{FF'} (-1)^{F+L'+J+J'+3/2} \sqrt{(2J+1)(2J'+1)} \times
\end{equation}
\[
\left\{
\begin{matrix}
L~2~L'\\
J'~F~J
\end{matrix}
\right\}
\left\{
\begin{matrix}
1~2~1\\
J~\frac{1}{2}~J'
\end{matrix}
\right\} (L||P^{(2)}||L'),
\]
\end{subequations}
where $\left\{
\begin{matrix}
a~b~c\\
d~e~f
\end{matrix}
\right\}$ are the $6j$-symbols defined in Ref.~\cite{varshalovich},
and
\begin{subequations}
\label{P2}
\begin{equation}
(L||P^{(2)}||L) = \sqrt{3} \hbar^2 \sqrt{\frac{L(2L+1)(2L+2)}{(2L-1)(2L+3)}} P_{L,L},
\end{equation}
\begin{equation}
(L-2||P^{(2)}||L) = -{3} \hbar^2 \sqrt{\frac{L(L-1)}{2L-1}} P_{L-2,L},
\end{equation}
\begin{equation}
(L+2||P^{(2)}||L) = -\frac{3}{2} \hbar^2 \sqrt{\frac{(2L+2)(2L+4)}{2L+3}} P_{L+2,L},
\end{equation}
\end{subequations}
with
\begin{equation}
\label{Pll}
P_{L,L} = \frac{d^2}{dr^2} + \frac{2}{r} \frac{d}{dr} - \frac{L(L+1)}{r^2}, \quad P_{L-2,L} = \frac{d^2}{dr^2} + \frac{2L+1}{r} \frac{d}{dr}+ \frac{L^2-1}{r^2},
\end{equation}
\[
P_{L+2,L} = \frac{d^2}{dr^2} - \frac{2L+1}{r} \frac{d}{dr} + \frac{L(L+2)}{r^2}.
\]
\end{widetext}

%
%
%

In the spherical approximation the $F$-exciton states correspond to
total momentum $F=5/2$ or $F=7/2$. Making use of Eqs.~\eqref{H} and
\eqref{Hamiltonian} we arrive at the following equations
\begin{widetext}
\begin{subequations}
\label{52}
\begin{equation}
\label{Psi52}
F= 5/2; \quad \Psi = g_{3,1/2}(\rho) |3,1/2,5/2\rangle + g_{1,3/2}(\rho) |1,3/2,5/2\rangle + g_{3,3/2}(\rho) |3,3/2,5/2\rangle,
\end{equation}
\begin{equation}
\label{H52}
\begin{pmatrix}
P_{3,3} + \frac{2}{r} & \sqrt{\frac{6}{5}}\mu P_{3,1} & -\frac{2}{\sqrt{5}}\mu P_{3,3}\\
\sqrt{\frac{6}{5}}\mu P_{1,3} & (1+\mu/5) P_{1,1} + \frac{2}{r} - \bar \Delta & -\frac{2\sqrt{6}}{{5}}\mu P_{1,3}\\
-\frac{2}{\sqrt{5}}\mu P_{3,3} & -\frac{2\sqrt{6}}{{5}}\mu P_{3,1} & (1-\mu/5) P_{3,3} + \frac{2}{r} - \bar \Delta
\end{pmatrix}
\begin{pmatrix}
g_{3,1/2}\\
g_{1,3/2}\\
g_{3,3/2}
\end{pmatrix} = E
\begin{pmatrix}
g_{3,1/2}\\
g_{1,3/2}\\
g_{3,3/2}
\end{pmatrix}.
\end{equation}
\end{subequations}

\begin{subequations}
\label{72}
\begin{equation}
\label{Psi72}
F= 7/2; \quad \Psi = g_{3,1/2}(\rho) |3,1/2,7/2\rangle + g_{3,3/2}(\rho) |3,3/2,7/2\rangle + g_{5,3/2}(\rho) |5,3/2,7/2\rangle,
\end{equation}
\begin{equation}
\label{H72}
\begin{pmatrix}
P_{3,3} + \frac{2}{r} & {\frac{1}{\sqrt{3}}}\mu P_{3,3} & -\sqrt{\frac{5}{3}}\mu P_{3,5}\\
{\frac{1}{\sqrt{3}}}\mu P_{3,3} & (1-2\mu/3) P_{3,3} + \frac{2}{r} - \bar \Delta & -\frac{\sqrt{5}}{{3}}\mu P_{3,5}\\
-\sqrt{\frac{5}{3}}\mu P_{5,3} & -\frac{\sqrt{5}}{{3}}\mu P_{5,3} & (1+2\mu/3) P_{5,5} + \frac{2}{r} - \bar \Delta
\end{pmatrix}
\begin{pmatrix}
g_{3,1/2}\\
g_{3,3/2}\\
g_{5,3/2}
\end{pmatrix} = E
\begin{pmatrix}
g_{3,1/2}\\
g_{3,3/2}\\
g_{5,3/2}
\end{pmatrix}.
\end{equation}
\end{subequations}
\end{widetext}


%

The calculated binding energies of the $P$-excitons and splittings
between $P$-and $F$-excitons are presented in Fig.~3(a) and (b) of
the main text. Our calculations show (see below for more details)
that the splitting between the $F=5/2$ and $F=7/2$ $F$-exciton
states is much smaller compared with the splitting between the $F$-
and $P$-excitons. Therefore the fine structure of $F$ excitons is
not shown in Fig. 3 of the main text.

\subsection{Effects of cubic symmetry}

We follow Refs.~\cite{PhysRevB.9.1525,Lipari1978665} and include
cubic contributions to the hole Hamiltonian in the form [cf. Eq. (2)
of the main text]
\begin{equation}
\label{H:cubic}
\mathcal H_c = \frac{\delta}{3\hbar^2} \left(\sum_{k=\pm 4}[P^{(2)} \times I^{(2)}]_k^{(4)}
+ \frac{\sqrt{70}}{5} [P^{(2)} \times I^{(2)}]_0^{(4)}\right),
\end{equation}
where $\delta = (\gamma_3 -\gamma_2)/\gamma_1'$. Generally, this
contribution mixes states with different $F_z$ for given $F$ and
with different $F$. This is because in cubic symmetry the angular
momentum is not a good quantum number. The matrix elements of
Eq.~\eqref{H:cubic} can be calculated in a way similar to the
derivation of Eq.~\eqref{Hmu} and Eq. (A2) in
Ref.~\cite{PhysRevB.9.1525}:
\begin{multline}
\label{me:delta}
\langle L'J'F'F_z'| \left[P^{(2)} \times \cdot I^{(2)} \right]^{(4)}_{m}|LJFF_z\rangle \\ = 9\sqrt{5}\times (-1)^{F'-F_z' +J+3/2} \times \\
\sqrt{(2J+1)(2J'+1)(2F+1)(2F'+1)} \times
\end{multline}
\[
\left(
\begin{matrix}
F' & 4 & F\\
-F_z' & m & F_z
\end{matrix}
\right)
\left\{
\begin{matrix}
J' & J & 2\\
L' & L & 2 \\
F' & F & 2
\end{matrix}
\right\}
\left\{
\begin{matrix}
1~2~1\\
J~\frac{1}{2}~J'
\end{matrix}
\right\} (L||P^{(2)}||L'),
\]
with the $3j$ symbols $\left(
\begin{matrix}
a & b & c\\
d & e & f
\end{matrix}
\right)$
 and $9j$ symbols $\left\{
\begin{matrix}
a & b & c\\
d & e & f \\
g & h & j
\end{matrix}
\right\}$ defined in Ref.~\cite{varshalovich}.

The wavefunctions of the states satisfy the following
equations~\footnote{Parts of these equations (for $F=5/2$ and
restricted to the $\Gamma_8^+$ representation) were derived in
Ref.~\cite{PhysRevB.9.1525} for the problem of an acceptor. Note
that there are two minor misprints in Ref.~\cite{PhysRevB.9.1525}:
in the last element of Eq.~\eqref{H52:8} they give $34/115$ instead
of $34/175$ and in the expression for one of the $9j$ symbols they
give the middle line as $1~1~1$ instead of $1~1~2$.}
\begin{widetext}
\begin{subequations}
\label{52:7}
\begin{equation}
\label{Psi52:7}
F= 5/2, \Gamma_6^-; \quad \Psi = g_{3,1/2}(\rho) |3,1/2,5/2, \Gamma_6^-\rangle + g_{1,3/2}(\rho) |1,3/2,5/2, \Gamma_6^-\rangle + g_{3,3/2}(\rho) |3,3/2,5/2,\Gamma_6^-\rangle,
\end{equation}
\begin{equation}
\label{H52:7}
\begin{pmatrix}
P_{3,3} + \frac{2}{r} & \left(\sqrt{\frac{6}{5}}\mu - \frac{8\delta}{35} \sqrt{\frac{6}{5}}  \right) P_{3,1} & \left(-\frac{2}{\sqrt{5}}\mu  +\frac{16\delta}{35\sqrt{5}}  \right) P_{3,3}\\
\left(\sqrt{\frac{6}{5}}\mu - \frac{8\delta}{35} \sqrt{\frac{6}{5}}  \right)P_{1,3} & (1+\frac{\mu}{5}+ \frac{24\delta }{25}) P_{1,1} + \frac{2}{r} - \bar \Delta & \left(-\frac{2\sqrt{6}}{{5}}\mu  + \frac{2\sqrt{6}}{25}\delta\right) P_{1,3}\\
\left(-\frac{2}{\sqrt{5}}\mu + \frac{16\delta}{35\sqrt{5}}\right)P_{3,3} & \left(-\frac{2\sqrt{6}}{{5}}\mu  + \frac{2\sqrt{6}}{25}\delta\right)P_{3,1} & (1-\frac{\mu}{5} - \frac{68\delta}{175}) P_{3,3} + \frac{2}{r} - \bar \Delta
\end{pmatrix}
\begin{pmatrix}
g_{3,1/2}\\
g_{1,3/2}\\
g_{3,3/2}
\end{pmatrix} = E
\begin{pmatrix}
g_{3,1/2}\\
g_{1,3/2}\\
g_{3,3/2}
\end{pmatrix}.
\end{equation}
\end{subequations}

\begin{subequations}
\label{52:8}
\begin{equation}
\label{Psi52:8}
F= 5/2, \Gamma_8^-; \quad \Psi = g_{3,1/2}(\rho) |3,1/2,5/2, \Gamma_8^-\rangle + g_{1,3/2}(\rho) |1,3/2,5/2, \Gamma_8^-\rangle + g_{3,3/2}(\rho) |3,3/2,5/2,\Gamma_8^-\rangle,
\end{equation}
\begin{equation}
\label{H52:8}
\begin{pmatrix}
P_{3,3} + \frac{2}{r} & \left(\sqrt{\frac{6}{5}}\mu + \frac{4\delta}{35} \sqrt{\frac{6}{5}}  \right) P_{3,1} & \left(-\frac{2}{\sqrt{5}}\mu  -\frac{8\delta}{35\sqrt{5}}  \right) P_{3,3}\\
\left(\sqrt{\frac{6}{5}}\mu + \frac{4\delta}{35} \sqrt{\frac{6}{5}}  \right)P_{1,3} & (1+\frac{\mu}{5}- \frac{12\delta }{25}) P_{1,1} + \frac{2}{r} - \bar \Delta & \left(-\frac{2\sqrt{6}}{{5}}\mu  - \frac{\sqrt{6}}{25}\delta\right) P_{1,3}\\
\left(-\frac{2}{\sqrt{5}}\mu - \frac{8\delta}{35\sqrt{5}}\right)P_{3,3} & \left(-\frac{2\sqrt{6}}{{5}}\mu  - \frac{\sqrt{6}}{25}\delta\right)P_{3,1} & (1-\frac{\mu}{5} + {\frac{34\delta}{175}}) P_{3,3} + \frac{2}{r} - \bar \Delta
\end{pmatrix}
\begin{pmatrix}
g_{3,1/2}\\
g_{1,3/2}\\
g_{3,3/2}
\end{pmatrix} = E
\begin{pmatrix}
g_{3,1/2}\\
g_{1,3/2}\\
g_{3,3/2}
\end{pmatrix}.
\end{equation}
\end{subequations}

\begin{subequations}
\label{72:A}
\begin{equation}
\label{Psi72:A}
F= 7/2, \Gamma_7^-; \quad \Psi = g_{3,1/2}(\rho) |3,1/2,7/2,\Gamma_7^-\rangle + g_{3,3/2}(\rho) |3,3/2,7/2,\Gamma_7^-\rangle + g_{5,3/2}(\rho) |5,3/2,7/2,\Gamma_7^-\rangle,
\end{equation}
\begin{equation}
\label{H72:A}
\begin{pmatrix}
P_{3,3} + \frac{2}{r} & \left({\frac{\mu}{\sqrt{3}}} + \frac{16\delta}{15\sqrt{3}} \right)P_{3,3} & -\left(\sqrt{\frac{5}{3}}\mu + \frac{8\delta}{33\sqrt{15}} \right) P_{3,5}\\
\left({\frac{\mu}{\sqrt{3}}} + \frac{16\delta}{15\sqrt{3}} \right) P_{3,3} & (1-\frac{2\mu}{3} - \frac{8\delta}{45}) P_{3,3} + \frac{2}{r} - \bar \Delta & -\left(\frac{\sqrt{5}}{{3}}\mu - \frac{28\delta}{99\sqrt{5}}\right) P_{3,5}\\
-\left(\sqrt{\frac{5}{3}}\mu + \frac{8\delta}{33\sqrt{15}} \right) P_{5,3} &  -\left(\frac{\sqrt{5}}{{3}}\mu - \frac{28\delta}{99\sqrt{5}}\right)P_{5,3} & (1+\frac{2\mu}{3} - \frac{56\delta}{495}) P_{5,5} + \frac{2}{r} - \bar \Delta
\end{pmatrix}
\begin{pmatrix}
g_{3,1/2}\\
g_{3,3/2}\\
g_{5,3/2}
\end{pmatrix} = E
\begin{pmatrix}
g_{3,1/2}\\
g_{3,3/2}\\
g_{5,3/2}
\end{pmatrix}.
\end{equation}
\end{subequations}

\begin{subequations}
\label{72:B}
\begin{equation}
\label{Psi72:B}
F= 7/2, \Gamma_6^-; \quad \Psi = g_{3,1/2}(\rho) |3,1/2,7/2,\Gamma_6^-\rangle + g_{3,3/2}(\rho) |3,3/2,7/2,\Gamma_6^-\rangle + g_{5,3/2}(\rho) |5,3/2,7/2,\Gamma_6^-\rangle,
\end{equation}
\begin{equation}
\label{H72:B}
\begin{pmatrix}
P_{3,3} + \frac{2}{r} & \left({\frac{\mu}{\sqrt{3}}} - \frac{16\sqrt{3}\delta}{35} \right)P_{3,3} & -\left(\sqrt{\frac{5}{3}}\mu - \frac{8\delta}{77} \sqrt{\frac{3}{5}} \right) P_{3,5}\\
\left({\frac{\mu}{\sqrt{3}}} - \frac{16\sqrt{3}\delta}{35} \right) P_{3,3} & (1-\frac{2\mu}{3} + \frac{8\delta}{35}) P_{3,3} + \frac{2}{r} - \bar \Delta & -\left(\frac{\sqrt{5}}{{3}}\mu + \frac{4\delta}{11\sqrt{5}}\right) P_{3,5}\\
 -\left(\sqrt{\frac{5}{3}}\mu - \frac{8\delta}{77} \sqrt{\frac{3}{5}} \right) P_{5,3} &  -\left(\frac{\sqrt{5}}{{3}}\mu + \frac{4\delta}{11\sqrt{5}}\right)P_{5,3} & (1+\frac{2\mu}{3} + \frac{8\delta}{55}) P_{5,5} + \frac{2}{r} - \bar \Delta
\end{pmatrix}
\begin{pmatrix}
g_{3,1/2}\\
g_{3,3/2}\\
g_{5,3/2}
\end{pmatrix} = E
\begin{pmatrix}
g_{3,1/2}\\
g_{3,3/2}\\
g_{5,3/2}
\end{pmatrix}.
\end{equation}
\end{subequations}

\begin{subequations}
\label{72:8}
\begin{equation}
\label{Psi72:8}
F= 7/2, \Gamma_8^-; \quad \Psi = g_{3,1/2}(\rho) |3,1/2,7/2,\Gamma_8^-\rangle + g_{3,3/2}(\rho) |3,3/2,7/2,\Gamma_8^-\rangle + g_{5,3/2}(\rho) |5,3/2,7/2,\Gamma_8^-\rangle,
\end{equation}
\begin{equation}
\label{H72:8}
\begin{pmatrix}
P_{3,3} + \frac{2}{r} & \left({\frac{\mu}{\sqrt{3}}} + \frac{16\delta}{105\sqrt{3}} \right)P_{3,3} & -\left(\sqrt{\frac{5}{3}}\mu + \frac{8\delta}{231\sqrt{15}} \right) P_{3,5}\\
\left({\frac{\mu}{\sqrt{3}}} + \frac{16\delta}{105\sqrt{3}} \right) P_{3,3} & (1-\frac{2\mu}{3} - \frac{8\delta}{315}) P_{3,3} + \frac{2}{r} - \bar \Delta & -\left(\frac{\sqrt{5}}{{3}}\mu - \frac{4\delta}{99\sqrt{5}}\right) P_{3,5}\\
-\left(\sqrt{\frac{5}{3}}\mu + \frac{8\delta}{231\sqrt{15}} \right)P_{5,3} &  -\left(\frac{\sqrt{5}}{{3}}\mu - \frac{4\delta}{99\sqrt{5}}\right)P_{5,3} & (1+\frac{2\mu}{3} - \frac{8\delta}{495}) P_{5,5} + \frac{2}{r} - \bar \Delta
\end{pmatrix}
\begin{pmatrix}
g_{3,1/2}\\
g_{3,3/2}\\
g_{5,3/2}
\end{pmatrix} = E
\begin{pmatrix}
g_{3,1/2}\\
g_{3,3/2}\\
g_{5,3/2}
\end{pmatrix}.
\end{equation}
\end{subequations}
\end{widetext}

Our calculations show that, in fact,  $F=5/2$ and $F=7/2$ are very
close to each other in energy. Therefore, we need to take into
account their mixing by the Hamiltonian~\eqref{H:cubic}. Using the
approach of Ref.~\cite{PhysRevB.9.1525} and Eq.~\eqref{me:delta} we
obtain the following effective Hamiltonians:
\begin{subequations}
\label{cub:mix}
\begin{equation}
\label{Gamma7:mixed}
\mathcal H^{(\Gamma_6^-)} =
\begin{pmatrix}
\mathcal H(5/2,\Gamma_6^-) & \mathcal X_6\\
\mathcal X_6^\dag & \mathcal H(7/2,\Gamma_6^-)
\end{pmatrix},
\end{equation}
where $\mathcal H(5/2,\Gamma_6^-)$ and $\mathcal H(7/2,\Gamma_6^-)$
are given by Eqs.~\eqref{H52:7} and \eqref{H72:B}, respectively, and
\begin{equation}
\label{X7}
\mathcal X_6 =
\begin{pmatrix}
{0} & {+}\frac{8\delta}{21}P_{3,3} &  {-}\frac{4\sqrt{5}\delta}{231}P_{3,5}\\
{-\frac{6\delta}{7}\sqrt{\frac{2}{5}}P_{1,3}} & 0 & 0\\
  {+}\frac{4\delta}{7}\sqrt{\frac{3}{ {5}}}P_{3, {3}} & {-}\frac{4\sqrt{5}\delta}{21}P_{3,3} & {+}\frac{4\delta}{33} P_{3,5}
\end{pmatrix},
\end{equation}
\end{subequations}
for the $\Gamma_6$ representation and
\begin{subequations}
\begin{equation}
\label{Gamma8:mixed}
\mathcal H^{(\Gamma_8^-)} =
\begin{pmatrix}
\mathcal H(5/2,\Gamma_8^-) & \mathcal X_8\\
\mathcal X_8^\dag & \mathcal H(7/2,\Gamma_8^-)
\end{pmatrix},
\end{equation}
where $\mathcal H(5/2,\Gamma_8^-)$ and $\mathcal H(7/2,\Gamma_8^-)$
are given by Eqs.~\eqref{H52:8} and \eqref{H72:8}, respectively, and
\begin{equation}
\label{X8}
\mathcal X_8 =
\begin{pmatrix}
{0} & {+}\frac{8\delta}{7\sqrt{15}}P_{3,3} & {-}\frac{4\delta}{77\sqrt{4}}P_{3,5}\\
{-\frac{6\sqrt{6}\delta}{35}P_{1,3}} & 0 & 0\\
{+}\frac{12\delta}{35}P_{3,3} & {-}\frac{4\delta}{7\sqrt{3}}P_{3,3} &{+}\frac{4\delta}{11\sqrt{15}} P_{3,5}
\end{pmatrix},
\end{equation}
\end{subequations}
for the $\Gamma_8^-$ representation.

%


\section{Transition energies and oscillator strengths}

\begin{figure}[h]
\includegraphics[width=0.8\linewidth]{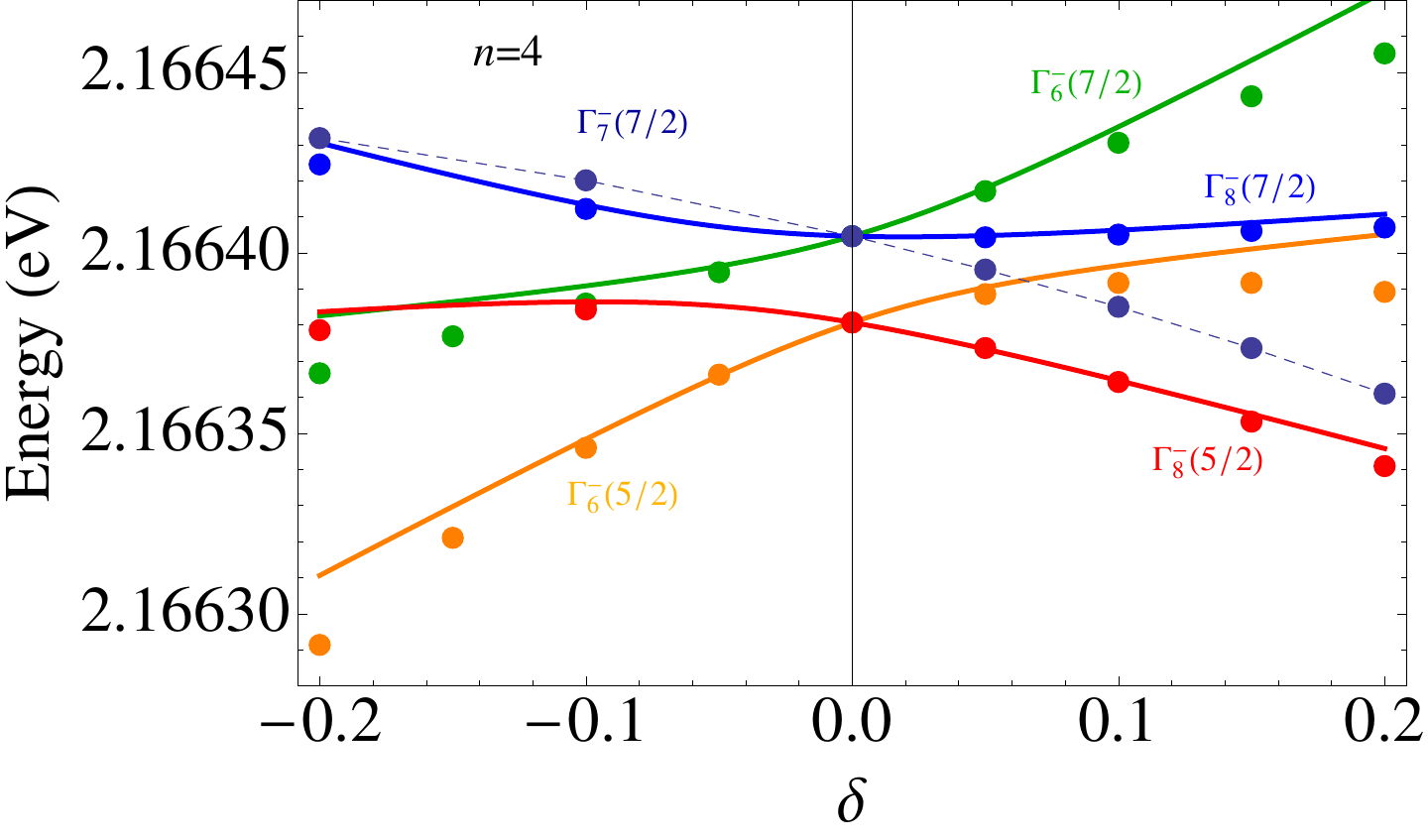}
\caption{Calculated energies of the $F$-excitons for the principle
quantum number $n=4$. Dots are the results of numerical
calculations, solid lines are the description in the framework of
the analytical model, Eq.~\eqref{analyt2}. States are labeled
according to the representations of the $O_h$ group for the hole
state Eq.~\eqref{hole:sum}, numbers in brackets denote the total
momentum $F$ of states for $\delta=0$. Note that the $\Gamma_7^-$
state is dark. The parameters of the calculation are: $\mathcal
R^*=87$~meV, $\mu=0.47$,
$\Delta=134$~meV~\cite{PhysRevB.23.2731}}\label{fig:delta}
\end{figure}

Figure~\ref{fig:delta} shows the energies of the $F$-excitons
calculated according to the developed model as function of the cubic
anisotropy parameter $\delta$. The other parameters were taken from
Ref.~\cite{PhysRevB.23.2731}. The dots give the numerical results,
the states are labeled according to the representation of the hole
state in Eq.~\eqref{hole:sum}. For $\delta=-0.1$ two of the four
$F$-exciton states become almost degenerate and, moreover, the
values of splittings between the non-degenerate states are close to
the experimental data (see also Fig. 3(c) of the main text). The
absolute values of the calculated transition energies deviate
slightly from the energies measured experimentally. This deviation
can be corrected by minor variation of the band gap energy in our
calculations.

\begin{figure}[h]
\includegraphics[width=0.8\linewidth]{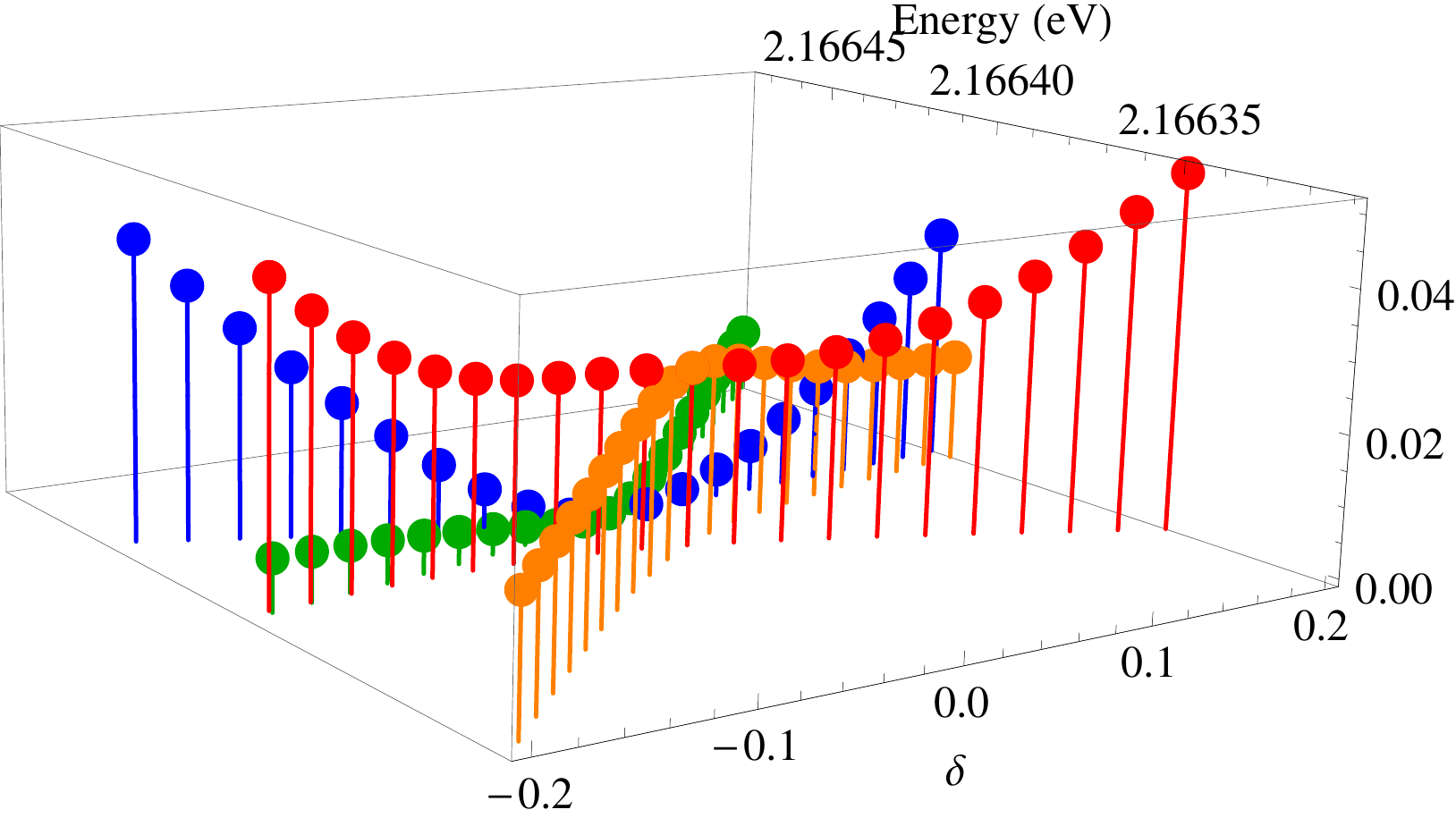}\\
\includegraphics[width=0.8\linewidth]{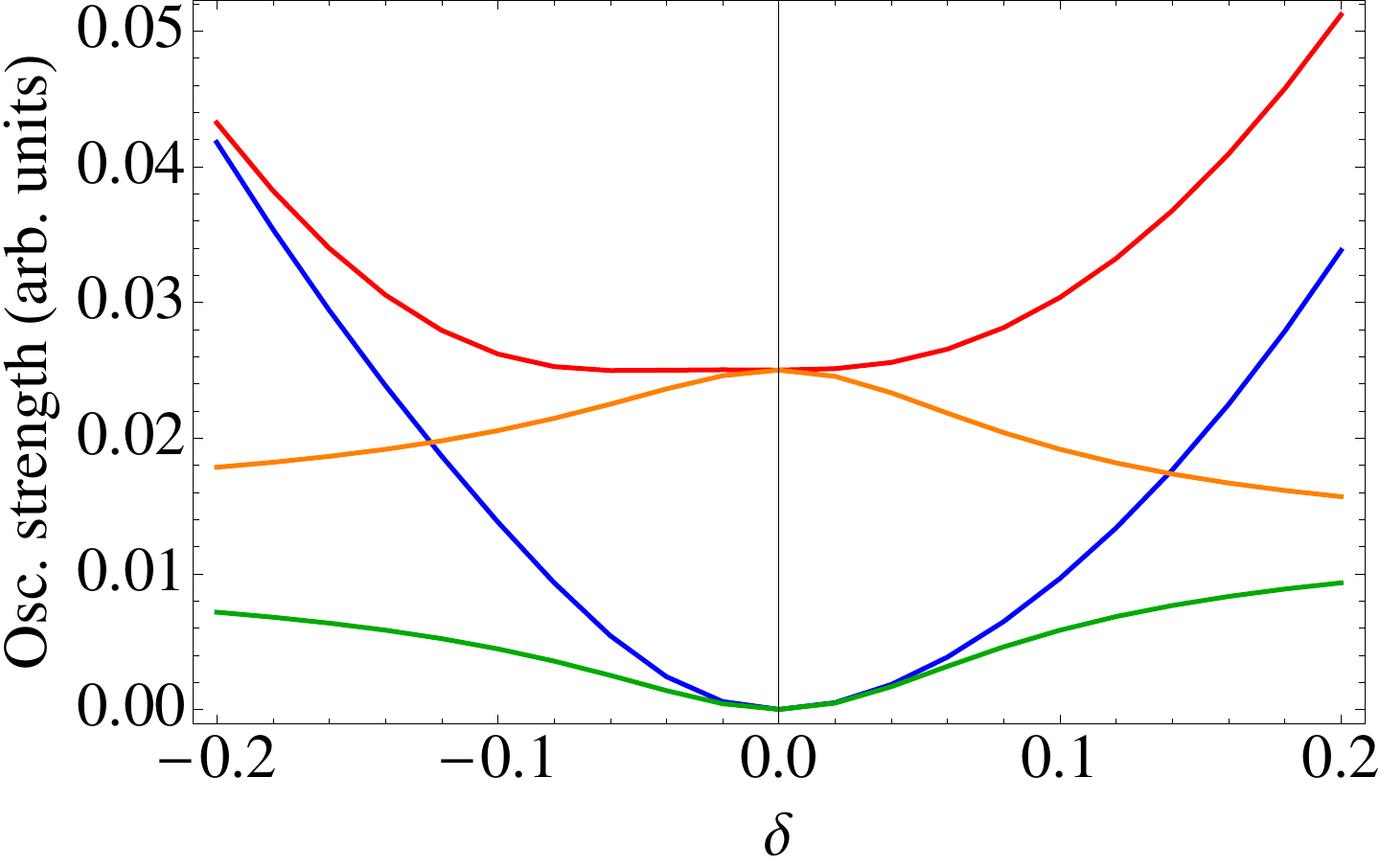}
\caption{Energies of $F$-shell states and their relative oscillator
strengths calculated in the framework of the two-level model,
Eqs.~\eqref{analyt2}. For the $\Gamma_8^-$ states two contributions
to the oscillator strength are taken into account, see text for
details.}\label{fig:osc}
\end{figure}

\begin{figure*}[!]
\includegraphics[width=0.45\linewidth]{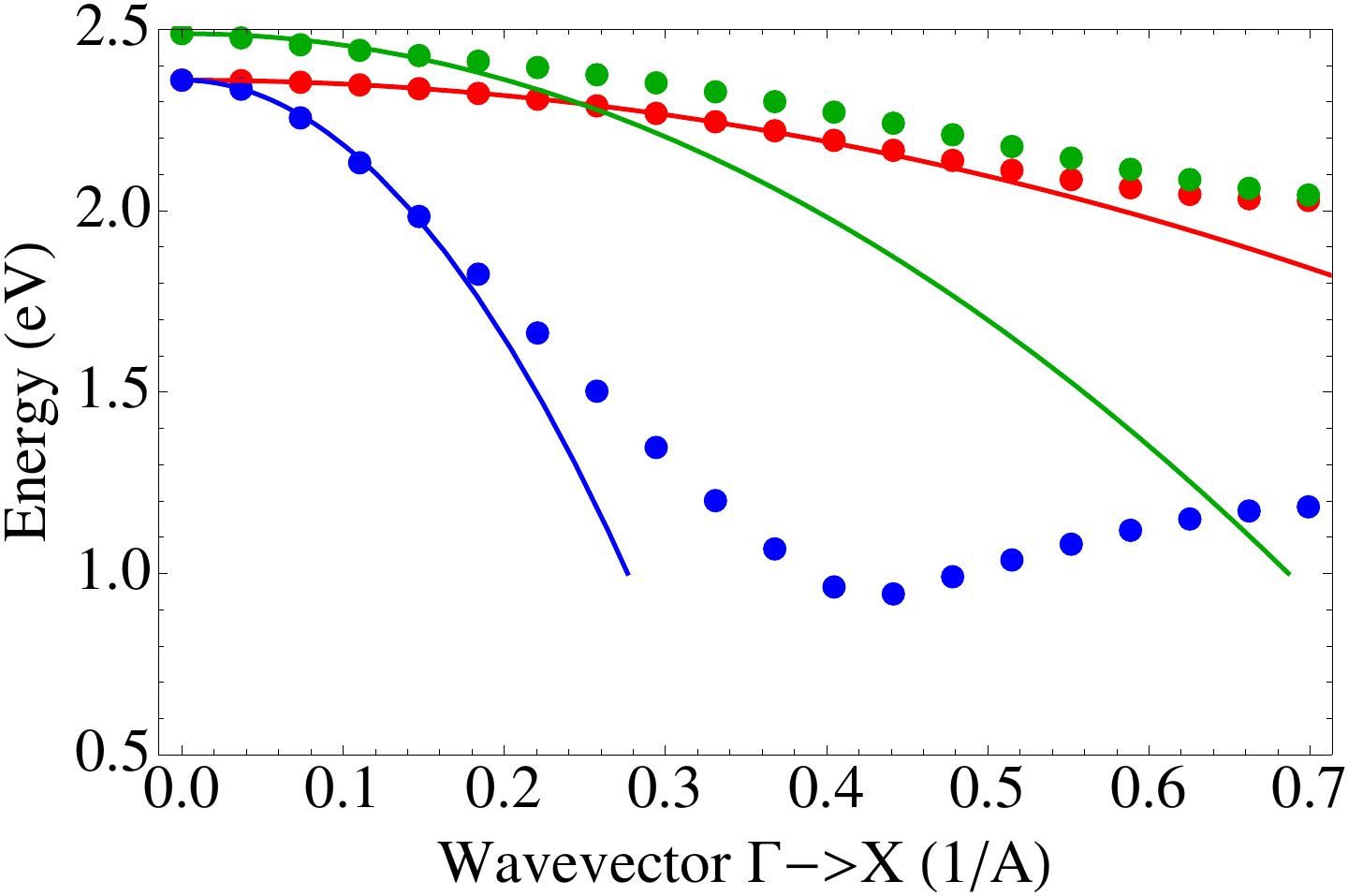} 
\includegraphics[width=0.45\linewidth]{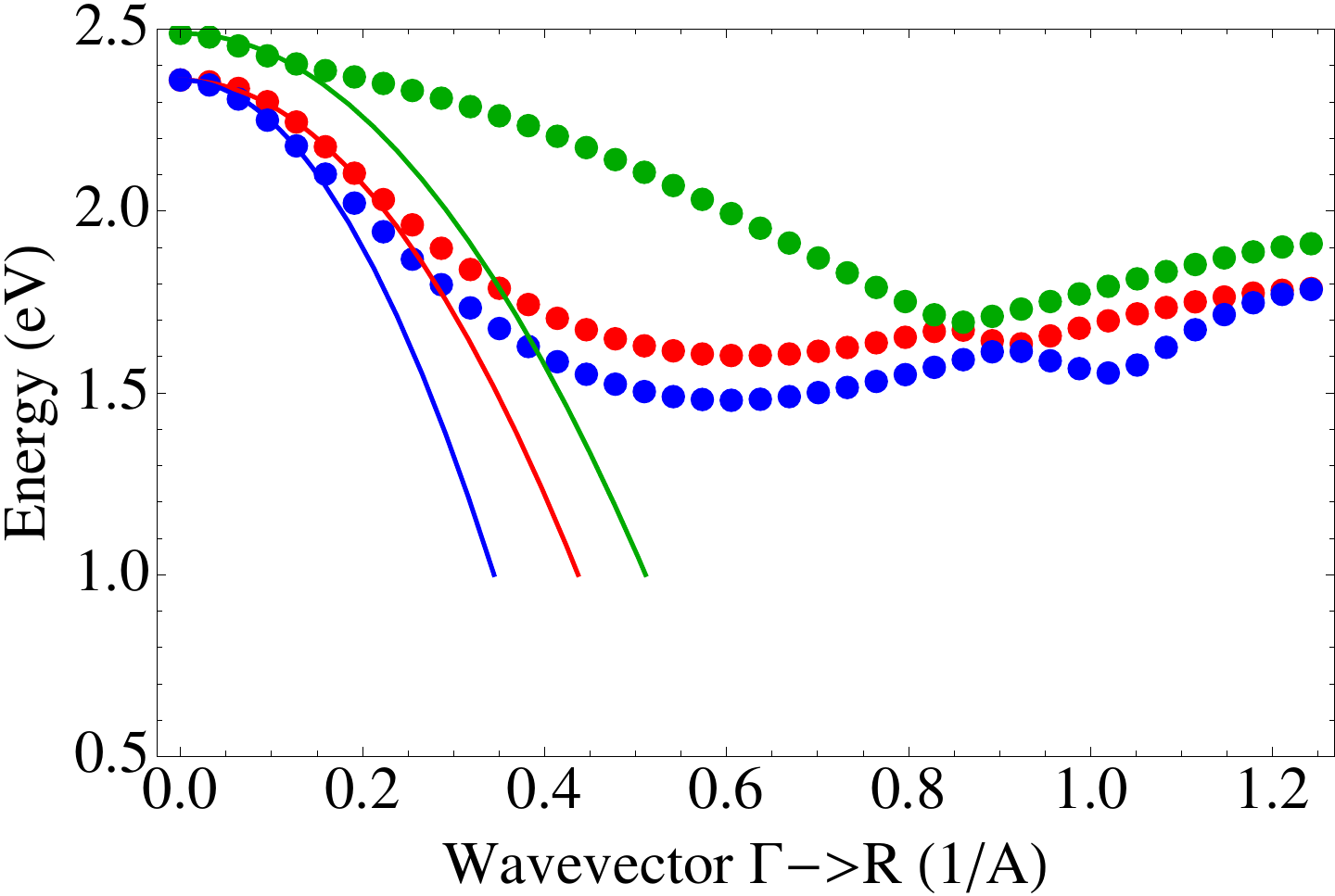}
\caption{Fit of the DFT band structure along the [001] and [111]
axes~\cite{0953-8984-21-1-015502}.}\label{fig:bands}
\end{figure*}

In order to obtain more insight into the level behavior we introduce
a two-level approximation valid for pairs of $\Gamma_6^-$ and
$\Gamma_8^-$ levels for a not too large cubic anisotropy parameter
$\delta$ as follows:
\begin{subequations}
\label{analyt2}
\begin{equation}
\mathcal H = \begin{pmatrix}
E_1 + \alpha_1 \delta & \beta \delta \\
\beta \delta & E_2 + \alpha_2 \delta,
\end{pmatrix}
\end{equation}
where $E_{1,2}$ are the energies of the states at $\delta=0$,
$\alpha_{1,2}$ describe the linear-with-$\delta$ shift of the
diagonal energies, $\beta$ describes the state mixing. The resulting
energies are
\begin{multline}
\label{energ2}
\mathcal E_{1,2} = \frac{E_1 + E_2 + (\alpha_1+ \alpha_2)\delta}{2} \\
\pm \sqrt{\left(\frac{E_1 - E_2 + (\alpha_1- \alpha_2)\delta}{2}\right)^2 + \beta^2\delta^2},
\end{multline}
\end{subequations}
and agree well with the numerical results, see the solid lines in
Fig.~\ref{fig:delta}.


To analyze the oscillator strengths we note that already in the
spherical approximation the states $\Gamma_8^-$ and $\Gamma_6^-$
originating from the $F=5/2$ level are optically active. This is
because the $F$-shell contribution with $\Gamma_7^+$ Bloch functions
is mixed with the $P$-shell envelope with $\Gamma_8^+$ Bloch
function, see Eq.~\eqref{Psi52}.

%

The cubic $O_h$ point symmetry of the cuprous oxide crystal results
in mixing of the $F=5/2$ and $F=7/2$ states described by
Eqs.~\eqref{cub:mix}. This makes optical transitions from the
$\Gamma_8^-$ and $\Gamma_6^-$ states allowed. This is the only
origin of optical activity for the $\Gamma_6^-(7/2)$ state shown by
the green points and lines in Figs.~\ref{fig:delta} and
\ref{fig:osc} (note also that the corresponding exciton wavefunction
contains substantial contribution from the $\Gamma_2^+$ Bloch
exciton state). For the $\Gamma_8^-$ states additional contributions
to the oscillator strength arise due to the presence of the $x^3$,
$y^3$, $z^3$-like envelopes of the $\Gamma_7^+$ Bloch functions
which, as a result of e.g. quartic contributions in the free-carrier
dispersion~\cite{ivchenko05a}
$$p_x^4+p_y^4+p_z^4,$$
contain $P$-shell contributions.  This contribution enhances the
oscillator strength of the $\Gamma_8^-(7/2)$ state as compared with
the $\Gamma_6^-(7/2)$ state.

\section{Determination of $\gamma_3$ sign from microscopic band structure calculations}

The analysis above, see Fig.~\ref{fig:delta} and Fig. 3 from the
main text shows that the best agreement of our calculations with
experiment is achieved for $\delta<0$, i.e. for $\gamma_3<\gamma_2$
(we recall that $\gamma_1$, $\gamma_2$ and $\gamma_3$ are the
Luttinger parameters). In order to verify this result we present in
Fig.~\ref{fig:bands}  the topmost valence band energies calculated
by DFT in Ref.~\cite{0953-8984-21-1-015502} and parabolic fits to
these results for the directions [001] (left) and [111] (right). The
accuracy of the fits is not enough to determine the values of all
Luttinger parameters but the sign of the difference $\gamma_3 -
\gamma_2$ can be reliably found. The estimation yields that
$\gamma_3 - \gamma_2$ is negative and $(\gamma_3-\gamma_2)/\gamma_1
\approx -0.3$ (for our parameters $(\gamma_3 - \gamma_2)/\gamma_1 =
\delta \gamma_1'/\gamma_1 \approx - 0.15$). The fact that
$\gamma_3<\gamma_2$ is also consistent with
Ref.~\cite{PhysRevB.56.7189}, where the DFT calculations were
carried out neglecting spin-orbit coupling. In the latter case,
however, the accuracy of extraction of parameters is low and a
direct calculation shows that in Ref.~\cite{PhysRevB.56.7189}
$\gamma_3$ is negative. We note that our values of $\gamma_1$,
$\gamma_2$ and $\gamma_3$ extracted from fitting the excitonic fine
structure may serve as benchmark for further improvement of
computational methods.


\end{document}